\documentclass[aps,twocolumn,groupedaddress,nofootinbib,floatfix]{revtex4-1}
\usepackage{graphicx}
\usepackage{color}
\usepackage{pifont}
\usepackage{amsmath,amsfonts,amssymb}
\usepackage{multirow}

\newcommand{\esx}{\langle S_1 \rangle}
\newcommand{\esy}{\langle S_2 \rangle}

\newcommand{\esz}{\langle S_3 \rangle}
\newcommand{\etz}{\langle T_0 \rangle}
\newcommand{\eax}{\langle A_1 \rangle}
\newcommand{\eay}{\langle A_2 \rangle}

\newcommand{\ebx}{\langle B_1 \rangle}
\newcommand{\eby}{\langle B_2 \rangle}

\newcommand{\oh}{\textstyle \frac{1}{2}}
\newcommand{\pmoh}{\textstyle\scriptstyle  \pm \frac{1}{2}}
\newcommand{\mpoh}{\textstyle\scriptstyle  \mp \frac{1}{2}}
\newcommand{\ot}{\textstyle \frac{1}{3}}

\newcommand{\ost}{\textstyle \frac{1}{\sqrt 2}}
\newcommand{\oss}{\textstyle \frac{1}{\sqrt 6}}

\newcommand{\soh}{{\textstyle \scriptstyle \frac{1}{2}}}
\newcommand{\smoh}{{\textstyle \scriptstyle - \frac{1}{2}}}

\newcommand{\smo}{{\textstyle \scriptscriptstyle -}1}

\DeclareMathOperator{\tr}{Tr}

\begin{document}

\title{Crafting polarisations for top, $W$ and $Z$}

\author{J. A. Aguilar-Saavedra}
\affiliation{Instituto de F\'\i sica Te\'orica, IFT-UAM/CSIC, c/ Nicolás Cabrera 13-15, 28049 Madrid}

\begin{abstract}
We put forward a method to tune the polarisation state of decaying heavy particles (top quarks and $W/Z$ bosons) in a pre-existing Monte Carlo sample. With this technique, dubbed as `custom angle replacement', the decay angular distributions are modified in such a way that the desired polarisation state is reproduced, while the production kinematics are unchanged. A non-trivial test of this approach is presented for the top quark semileptonic decay $t \to W b \to \ell \nu b$, with $\ell = e,\mu$, in which the decay distribution is four-dimensional and involves three different Lorentz frames. The proposed method can be used to obtain event samples with polarised heavy particles, as required by experimental measurements of polarisation and spin correlations. 
\end{abstract}

\maketitle

\section{Introduction}

Polarisation measurements provide a quite useful handle to obtain information about heavy particles and investigate their properties, in the search for new physics beyond the standard model (SM). The polarisation of the top quarks has been measured by the ATLAS and CMS Collaborations in pair~\cite{Aaboud:2019hwz,CMS:2019nrx} and single production~\cite{CMS:2015cyp,ATLAS:2022vym}. 
The polarisation of $W$ bosons produced in top decays has also been measured~\cite{CMS:2014uod,CMS:2016asd,ATLAS:2016fbc,ATLAS:2017ygi}, as well as the (joint) polarisation of the weak bosons in $WZ$ production~\cite{CMS:2021icx}. 
These delicate measurements are possible owing to the large statistics provided at the Large Hadron Collider (LHC). Polarisation measurements will continue to have a key role in the investigation of particle properties at the LHC Run 3, as well as at the high-luminosity LHC upgrade. However, despite the importance of these measurements, general tools are not available that allow to generate Monte Carlo samples where the polarisation of heavy particles can be chosen --- instead, it is determined by angular momentum conservation and the dynamics of the process under consideration.

Polarisation measurements are quite demanding. The polarisation of short-lived particles such as the top quark and the $W,Z$ bosons can be extracted from the angular distributions of their decay products. The parton-level distributions, which carry the imprint of the polarisation of the parent particle, are modified by the detector acceptance and resolution, as well as by kinematical cuts that have to be imposed in order to reduce backgrounds. Therefore, the parton-level distributions are not directly accesible. One technique often used to recover them is an unfolding: to reverse the detector effects to obtain the original parton-level quantities, which are subsequently analysed to determine the polarisation observables. Another possibility is to use templates: samples with definite polarisations (the so-called templates) are simulated and the measured data is fit with a combination of templates, thus obtaining the polarisation observables. Monte Carlo samples with definite polarisations are very useful in either case. In the former, they can be used to test the robustness of the unfolding. In the latter, they constitute the essential ingredient to build the templates. In this regard, we point out that template methods have been used by the ATLAS and CMS Collaborations to measure the polarisation of $W$ bosons~\cite{CMS:2014uod,CMS:2016asd,ATLAS:2016fbc} and top quarks~\cite{ATLAS:2022vym}. In addition, a template method has recently been proposed for the measurement of $t \bar t$ spin correlations in Ref.~\cite{Aguilar-Saavedra:2021ngj}. 

On the other hand, some interesting processes do not allow for polarisation measurements, because the full kinematical reconstruction of the momenta of decaying particles is not possible. In these cases it is still possible to obtain some information about spin by comparing the SM with alternative hypotheses. One such example is the Higgs boson decay $H \to W W \to \ell^+ \nu \ell^- \nu$.  Polarised $H \to WW$ samples may be used to obtain information about spin in this process, using laboratory (lab) frame observables~\cite{Aguilar-Saavedra:2022mpg}.
Furthermore, polarised samples may be useful to train polarisation-agnostic multivariate discriminants, e.g. neural networks, to reduce a possible bias in polarisation measurements caused by the multivariate tools.

To the best of our knowledge, 
two techniques have been used in the literature to build (pseudo-)polarised event samples:\footnote{After the completion of the first version of this manuscript we learned that the ATLAS Collaboration used in Ref.~\cite{ATLAS:2022vym} a method PolManip~\cite{polmanip} similar to ours, in order to modify the top decays in SM samples. As shown in section~\ref{sec:4} and Ref.~\cite{Aguilar-Saavedra:2022jzo}, polarised samples used as templates must include polarisation effects also in the production, if any, and the specific method used in Ref.~\cite{ATLAS:2022vym} deserves further scrutiny.}

\begin{itemize}
\item[(i)] An event reweighting based on the polar angle $\cos \theta^*$ charged lepton distribution (see the next section for details), has been used for $W$ polarisation measurements in top quark decays~\cite{CMS:2014uod,CMS:2016asd,ATLAS:2016fbc}. In general a reweighting is unsound because the correlation between the distribution considered and other variables, namely the azimuthal angle $\phi^*$, may induce a bias in the measurement.\footnote{For example, there is a strong correlation between these two variables for $W$ bosons resulting from the decay of polarised top quarks.}
\item[(ii)] Spin projectors for top quarks have been implemented at the matrix-element level in Ref.~\cite{Aguilar-Saavedra:2021ngj}. This procedure is fully consistent with the definition of polarisation for an unstable intermediate resonance~\cite{Aguilar-Saavedra:2022jzo}.
\end{itemize}
As the statistical precision of the measurement increases at the LHC Run 3 and beyond, it is compulsory to improve the modeling of polarised samples as far as possible. 
The technique put forward in this paper, dubbed as `custom angle replacement' (CAR), is based on modifying the decay angular distributions of a pre-existing sample according to the desired polarisation state, while keeping the production kinematics identical. Interestingly, it may be used in samples generated beyond the leading order (LO), as long as the decay of the heavy particles (top, $W$, $Z$) is considered at the LO, which is often a good approximation. We note that the production $\times$ decay factorisation breaks down if higher-order non-factorisable corrections are important, especially with respect to the experimental uncertainties. (In such case, the very definition of `polarisation' of a heavy particle may become ambiguous, for example if these corrections involve diagrams where this heavy particle is absent.) Therefore, the applicability of the CAR method depends on the level of theoretical precision required to match the experimental uncertainties.

The CAR method is outlined in section~\ref{sec:2}, providing the analytic expressions necessary for its implementation with LO decays of the top quark and $W/Z$ bosons. A test for the top quark decay demonstrating the robustness of the approach is presented in section~\ref{sec:3}. 
By design, the CAR method keeps the production kinematics of the pre-existing sample. However, as we have mentioned, top quark polarisation measurements using a template method~\cite{Aguilar-Saavedra:2021ngj} require that the polarised samples are generated with spin projectors at the matrix-element level --- so that the top production kinematics corresponds to that of polarised top quarks.
In such case the CAR method still proves to be useful, because it can be used to {\em augment} the size of a small polarised dataset to obtain a much larger sample, with events that can be regarded as statistically independent. By doing so, the CAR method provides a computational advantage, for example for event generation beyond the LO, or in processes with multiple top quarks. We will show how the samples augmented with the CAR method are well suited for template measurements in section~\ref{sec:4}, specifically considering the measurement of the combined polarisation of $t \bar t$ pairs produced at the LHC.  We discuss our results and possible extensions in section~\ref{sec:5}.

\section{The CAR method}
\label{sec:2}

Let us consider a decaying particle $t$, $W$ or $Z$. Its spin state, which is given by an appropriate density operator, fully determines the multi-dimensional angular distribution of its decay products.
Therefore, if we are able to modify this angular distribution in a Monte Carlo event sample, we can craft a new sample mimicking the desired polarisation of the decaying particle. This is the basic idea that underlies the CAR method. 

Using the Jacob-Wick helicity formalism~\cite{Jacob:1959at} it can be shown on general grounds that
for weak boson decays $W \to \ell \nu$, $Z \to \ell^+ \ell^-$, the decay distribution involves the two angles $(\theta^*,\phi^*)$ that describe the orientation of the leptons in the rest frame of $V=W,Z$. For top quarks $t \to W b \to \ell \nu b$, two angles $(\theta,\phi)$ 
determine the orientation of the $W$ boson and $b$ quark in the top rest frame, and two additional angles $(\theta^*,\phi^*)$ describe the orientation of the leptons in the $W$ rest frame.
Then the procedure to obtain a sample with a definite polarisation state consists in {\em replacing} event by event these angles by new ones, randomly generated according to a probability density function (p.d.f.) appropriate to the spin configuration desired, and re-computing the four-momenta from these new angles. Obviously, angular momentum is preserved by construction, because the p.d.f.'s themselves are obtained using angular momentum conservation.

In the remainder of this section we discuss in turn the case of weak bosons and the top quark. We first set up the notation and collect the decay angular distributions from Refs.~\cite{Aguilar-Saavedra:2015yza,Aguilar-Saavedra:2017zkn,Aguilar-Saavedra:2017wpl}, that are used as p.d.f. Then, we explicitly describe the procedure followed to replace the angles, which has many subtleties arising from the use of different Lorentz frames at the same time.

\subsection{$W$ and $Z$ bosons}
\label{sec:2.1}

We follow Refs.~\cite{Aguilar-Saavedra:2015yza,Aguilar-Saavedra:2017zkn} to describe the angular distribution for the decay of a weak boson $V=W,Z$ in an arbitrary spin state. By fixing a reference system $(x,y,z)$ in the $V$ rest frame, the density operator describing the spin state can be parameterised in terms of irreducible tensor operators of ranks 1 and 2. Let us define, as usual, the spin operators in the spherical basis
\begin{equation}
S_{\pm 1} = \mp \ost (S_1 \pm i S_2) \,,\quad S_0 = S_3 \,,
\end{equation}
 and five rank 2 irreducible tensors $T_M$, built from $S_M$ as
\begin{align}
& T_{\pm 2} = S_{\pm1}^2 \,, \quad T_{\pm 1} = \ost \left[ S_{\pm 1} S_0 + S_0 S_{\pm 1} \right] \,, \notag \\
& T_{0} = \oss \left[ S_{+1} S_{-1} + S_{-1} S_{+1} + 2 S_0^2 \right] \,.
\end{align} 
Then, the spin density operator can be written as\footnote{This expansion follows from the fact that $S_M$ and $T_M$ are linearly independent and traceless. The coefficients can be determined by computing the expected value of spin operators for an arbitrary linear combination of $S_M$ and $T_M$.}
\begin{equation}
\rho = {\ot} \openone + \oh \displaystyle \sum_{M=-1}^1 \langle S_M \rangle^* S_M + \displaystyle \sum_{M=-2}^2  \langle T_{M} \rangle^* T_{M} \,.
\label{ec:rhoV}
\end{equation}
For the purpose here, it is convenient to use Hermitian operators for the parameterisation, by defining
\begin{align}
A_1 = \frac{1}{2} (T_1-T_{-1}) \,, \quad A_2 = {\frac{1}{2 i}} (T_1 + T_{-1}) \,, \notag \\
B_1 = \frac{1}{2} (T_2+T_{-2}) \,, \quad B_2 = {\frac{1}{2 i}} (T_2 - T_{-2}) \,.
\end{align}
In terms of these, the matrix elements of $\rho$ in the basis of $S_3$ eigenstates $|+\rangle$, $|0\rangle$, $|-\rangle$ are given by
\begin{align}
	& \rho_{\pm 1 \pm 1} = \frac{1}{3} \pm \frac{1}{2} \esz + \frac{1}{\sqrt 6} \etz \,,  \notag \\
	& \rho_{\pm 10} = \frac{1}{2\sqrt 2} \left[ \esx \mp i \esy \right] \mp \frac{1}{\sqrt 2} \left[ \eax \mp i \eay \right]  \,, \notag \\
	& \rho_{00} = \frac{1}{3} - \frac{2}{\sqrt 6} \etz \,, \notag \\
	& \rho_{1\, -1} = \ebx - i \eby \,,
\label{ec:Mrhov}
\end{align}
with $\rho_{m'm}=\rho_{mm'}^*$. The eigenvalues of $A_i$ and $B_i$ are $-1/2,0,1/2$, therefore their expected values range between $-1/2$ and $1/2$. The eigenvalues of $T_0$ are $-\sqrt{2/3}$ and $1/6$, therefore its expected value is in the interval $[-\sqrt{2/3},1/6]$. The semi-positivity of $\rho$ implies that its eigenvalues are non-negative, but the resulting conditions are quite cumbersome to write analytically. On the other hand, the condition $\tr \rho^2 \leq 1$ implies
\begin{align}
& \frac{1}{2} \left[ \esx^2 + \esy^2 + \esz^2 \right] + \etz^2 \notag \\
& + 2 \left[ \eax^2 + \eay^2 + \ebx^2 + \eby^2 \right] \leq \frac{2}{3} \,.
\end{align}

For $W$ bosons there is only one decay amplitude once we take the leptons massless, because of the left-handed chirality of the $W$ coupling. For $Z$ bosons there are two amplitudes related by the ratio of the left- and right-handed couplings $g_R^\ell \,:\, g_L^\ell$. Consequently, the decay angular distribution has a slightly different form for $W^+$, $W^-$ and $Z$ bosons. 
For the former, we define $(\theta^*,\phi^*)$ as the polar and azimuthal angles of the charged lepton three-momentum ($\ell^+$ for $W^+$ and $\ell^-$ for $W^-$). For $Z$ bosons, we use the negative lepton. (In all cases, the momenta are taken in the $V$ rest frame.) Then, the differential distribution reads
\begin{align}
& \frac{1}{\Gamma} \frac{d\Gamma}{d\!\cos\theta^* d\phi^*} = \frac{3}{8\pi} \left\{ 
\frac{1}{2} (1+\cos^2 \theta^*) - \eta_\ell \esz \cos \theta^*
\right. \notag \\
& ~ + \left[ \frac{1}{6} - \frac{1}{\sqrt 6} \etz \right] \left( 1-3\cos^2 \theta^* \right) \notag \\
& ~ - \eta_\ell \esx \cos \phi^* \sin \theta^* - \eta_\ell \esy \sin \phi^* \sin \theta^* \notag \\ 
& ~ - \eax \cos \phi^* \sin 2\theta^* - \eay \sin \phi^* \sin 2\theta^* \notag \\
& \left. ~ + \ebx \cos 2 \phi^* \sin^2 \theta^* + \eby \sin 2 \phi^* \sin^2 \theta^* \right\} \,,
\label{ec:distV}
\end{align} 
where $\eta_\ell = -1$ for $W^+$, $\eta_\ell = +1$ for $W^-$ and 
\begin{equation}
\eta_\ell = \frac{(g_L^\ell)^2 - (g_R^\ell)^2}{(g_L^\ell)^2 + (g_R^\ell)^2} = \frac{1-4 s_W^2}{1-4 s_W^2 + 8 s_W^4} \simeq 0.13
\label{ec:etal}
\end{equation}
for $Z$ bosons, with $s_W$ the sine of the weak mixing angle. 

Using (\ref{ec:distV}) as p.d.f., one can craft a sample of $V$ bosons in an arbitrary spin state, given by a physical density operator $\rho$. The procedure to be followed, event by event, is:
\begin{enumerate}
\item We take the four-momenta of the decay products (leptons) in the lab frame, which we label as $p_{L_1}$ and $p_{L_2}$. In the case of $W$ decays $L_1$ is the charged lepton and $L_2$ the neutrino, while for $Z$ decays $L_1$ is the negative lepton.
\item The energy and modulus of the three-momentum of both leptons are computed in the $V$ rest frame. We label these quantities as $E_{L_1}$, $|\vec p_{L_1}|$, $E_{L_2}$, $|\vec p_{L_2}|$. Because these are rotationally-invariant quantities, the precise way in which the boosts are performed is unimportant.
\item We generate $\cos \theta^*$, $\phi^*$ according to the p.d.f. in (\ref{ec:distV}), for the values of $\langle S_i \rangle$, $\langle A_i \rangle$, $\langle B_i \rangle$ and $\langle T_0 \rangle$ that correspond to the specific density operator $\rho$ under consideration. To this end, one can use for example the acceptance-rejection method.
\item We define the new $L_1$ four-momentum in the $V$ rest frame $p_{L_1}^{\prime R}$, with energy 
$E_{L_1}$ and three-momentum in the $(\theta^*,\phi^*)$ direction and modulus $|\vec p_{L_1}|$. For the $L_2$ momentum $p_{L_2}^{\prime R}$ we do similarly, but in the opposite spatial direction. 
\item The new momenta $p_{L_1}^\prime$, $p_{L_2}^\prime$ in the lab frame are obtained by a two-step boost: first a pure boost of $p_{L_1}^{\prime R}$, $p_{L_2}^{\prime R}$ from the $V$ rest frame to the centre-of-mass (c.m.) frame, followed by a pure boost from the c.m. frame to the lab frame.
\end{enumerate}
The four-momenta of $V$, as well as of any other particles, are not changed, therefore the kinematics of the production is maintained.

\subsection{Top quarks}
\label{sec:2.2}

We follow Ref.~\cite{Aguilar-Saavedra:2017wpl} to parameterise the fully-differential distribution for the cascade decay $t \to W b \to \ell \nu b$.  Let us fix a reference system $(x,y,z)$ in the rest frame of the top quark. (We later point out the differences for anti-quarks.) It is customary to parameterise the density operator as
\begin{equation}
\rho = \frac{1}{2} (\openone + \vec P \cdot \vec \sigma)  \,, 
\label{ec:rhoT}
\end{equation}
with $\sigma$ the Pauli matrices, and $\vec P$ the so-called polarisation vector, with components $P_i \equiv 2 \langle S_i \rangle$. In contrast with the case of massive spin-1 bosons, the density operator is parameterised with only three quantities, related to the expected values of the spin operators. In the basis of $S_3$ eigenstates $|+\rangle$, $|-\rangle$, the matrix elements are
\begin{align}
& \rho_{\pmoh \, \pmoh} = \frac{1}{2} (1 \pm P_3) \,, \quad \notag \\
& \rho_{\pmoh \, \mpoh} = \frac{1}{2} (P_1 \mp i P_2 ) \,.
\end{align}
The density operator $\rho$ defined by the above parameterisation is physical as long as $|\vec P| \leq 1$. 

There are only four decay amplitudes for $t \to W b$, because the total angular momentum in the $W$ flight direction in the top rest frame has to be $\pm 1/2$. Labelling them as $a_{\lambda_1 \lambda_2}$, with $\lambda_1,\lambda_2$ the helicities of the $W$ boson and $b$ quark, respectively, the non-zero amplitudes are $a_{1 \,\soh}$, $a_{0 \, \smoh}$, $a_{0 \, \soh}$ and $a_{\smo \, -\soh}$. 

The fully-differential top decay distribution can be parameterised with four angles: $(\theta,\phi)$ are the polar coordinates of the $W$ boson three-momentum in the top rest frame, and $(\theta^*,\phi^*)$ are the polar coordinates of the charged lepton three-momentum in the $W$ rest frame. The reference system for $(\theta,\phi)$ is the same one $(x,y,z)$ used to express the density operator $\rho$. The orientation of the reference system $(x',y',z')$ for the $W$ rest frame results from the definition of helicity states (see for example Ref.~\cite{libro}): 
\begin{itemize}
\item The $\hat z'$ axis is in the direction of the $W$ boson three-momentum in the top rest frame, $\hat z' = \sin \theta \cos \phi \, \hat x + \sin \theta \sin \phi \, \hat y + \cos \theta \, \hat z$
\item The $\hat y'$ axis is in the $xy$ plane, making an angle $\phi$ with the $\hat y$ axis: $\hat y' = - \sin \phi \, \hat x + \cos \phi \, \hat y$.
\item The $\hat x'$ axis is orthogonal to both, $\hat x' = \hat y' \times \hat z' = \cos \theta \cos \phi \, \hat x + \cos \theta \sin \phi \, \hat y - \sin \theta \hat z$.  
\end{itemize}

For convenience in the notation, we define the sum of squared amplitudes
\begin{equation}
\mathcal{N} = |a_{1 \, \soh}|^2 + |a_{0 \, \soh}|^2 + |a_{0 \, \smoh}|^2 + |a_{\smo \, \smoh}|^2  \,.
\end{equation}
With these conventions, and assuming that the amplitudes do not have a relative complex phase (as it happens in the SM), the differential distribution reads
\begin{align}
& \frac{1}{\Gamma} \frac{d\Gamma}{d\Omega d\Omega^*} \notag \\
& \quad = \frac{3}{64\pi^2} \frac{1}{\mathcal{N}}
\left\{   \left[ |a_{1 \,\soh}|^2 \left( 1+\lambda \cos \theta^* \right)^2 + 2 |a_{0 \, \smoh}|^2 \sin^2 \theta^*  \right] \right.  \notag \\
& \quad \times \left( 1 + \vec P \cdot \hat z' \right) \notag \\
& \quad + \left[ 2 |a_{0 \, \soh}|^2 \sin^2 \theta^*  + |a_{\smo \, \smoh}|^2 \left( 1-\lambda \cos \theta^* \right)^2 \right] \notag \\
& \quad \times \left( 1 - \vec P \cdot \vec \hat z' \right) \notag \\
%
& \quad + \lambda 2 \sqrt 2 \left[ a_{0 \, \soh} a_{1 \, \soh}^* ( 1+\lambda \cos \theta^*) \right. \notag \\
& \quad \left. + a_{\smo \, \smoh} a_{0 \, \smoh}^*  (1-\lambda \cos \theta^*)
 \right] \cos \phi^* \sin \theta^* \vec P \cdot \hat x' \notag \\
%
&\quad + \lambda 2 \sqrt 2  \left[ a_{0 \, \soh} a_{1 \, \soh}^* (1+\lambda \cos \theta^*) \right. \notag \\
& \quad + \left. \left. a_{\smo \, \smoh} a_{0 \, \smoh}^*  (1-\lambda \cos \theta^*) 
\right] \sin \phi^* \sin \theta^* \vec P \cdot \hat y' \right\} \,,
\label{ec:distT}
\end{align}
with $\lambda = 1$ for top quarks and $\lambda=-1$ for anti-quarks. At LO, the amplitude products entering the differential distribution are 
\begin{align}
& |a_{1 \, \soh}|^2 = \left( 1 - \frac{M_W^2}{m_t^2} \right) - 2 \frac{|\vec p_W|}{m_t}  \,, \notag \\
&  |a_{\smo \, \smoh}|^2 = \left( 1 - \frac{M_W^2}{m_t^2} \right) + 2 \frac{|\vec p_W|}{m_t} \,, \notag \\ 
& |a_{0 \, \soh}|^2 = \frac{1}{2}  \left( \frac{m_t^2}{M_W^2} - 1 \right) - \frac{|\vec p_W| m_t}{M_W^2} \,, \notag \\
&   |a_{0 \, \smoh}|^2 = \frac{1}{2}  \left( \frac{m_t^2}{M_W^2} - 1 \right) + \frac{|\vec p_W| m_t}{M_W^2} \,, \notag \\
& a_{0 \, \soh} a_{1 \, \soh}^* = \frac{m_t}{\sqrt 2 M_W} \left( 1 - \frac{M_W^2}{m_t^2} \right) - \sqrt{2}  \frac{|\vec p_W|}{M_W}  \,, \notag \\
& a_{0 \, \smoh} a_{\smo \, \smoh}^* = \frac{m_t}{\sqrt 2 M_W} \left( 1 - \frac{M_W^2}{m_t^2} \right) + \sqrt{2}  \frac{|\vec p_W|}{M_W} 
\label{ec:amp}
\end{align}
for top quarks, with $m_t$ and $M_W$ the top quark and $W$ boson masses, and $|\vec p_W|$ the modulus of the $W$ boson three-momentum in the top rest frame. For anti-quarks, the decay amplitudes (denoted below by a bar) are related to the top quark ones by
\begin{align}
& | \bar a_{1 \, \oh}|^2 = |a_{\smo \, \smoh}|^2 \,, \quad \quad  |\bar a_{\smo \, \smoh}|^2 = |a_{1 \, \oh}|^2 \,, \notag \\
& |\bar a_{0 \, \oh}|^2 = |a_{0 \, \smoh}|^2 \,, \quad \quad |\bar a_{0 \, \smoh}|^2 = | a_{0 \, \oh}|^2 \,, \notag \\
& \bar a_{0 \, \oh} \bar a_{1 \, \oh}^* = \left( a_{0 \, \smoh} a_{\smo \, \smoh}^* \right)^* \,, \notag \\
& \bar a_{0 \, \smoh} \bar a_{\smo \, \smoh}^* = \left( a_{0 \, \oh} a_{1 \, \oh}^* \right)^* \,.
\end{align}
Using (\ref{ec:distT}) as p.d.f., one can apply the CAR method to craft polarised top samples. The procedure is more involved than for weak bosons because of the intermediate decay, but we detail it below for reference and reproducibility. Event by event, the flow is:
\begin{enumerate}
\item We take the four-momenta of the $W$ boson $p_W$, $b$ quark $p_b$ and leptons $p_{\ell}$, $p_{\nu}$ in the lab frame.
\item The energy and the modulus of the three-momentum of $W$ and $b$ are computed in the top rest frame, and likewise the energy and modulus of the three-momentum of the leptons in the $W$ rest frame. We label these quantities as $E_W$, $|\vec p_W|$, $E_b$, $|\vec p_b|$, $E_{\ell}$, $|\vec p_{\ell}|$, $E_{\nu}$, $|\vec p_{\nu}|$ in obvious notation. Because these are rotationally-invariant quantities, the precise way in which the boosts are performed is unimportant.
\item We generate $\cos \theta$, $\phi$, $\cos \theta^*$, $\phi^*$ according to the p.d.f. in (\ref{ec:distT}), for the values of $P_i$ that correspond to the specific density operator $\rho$ under consideration.
\item The new $W$ boson four-momentum in the top rest frame $p_W^{\prime R}$ is defined with energy 
$E_W$ and three-momentum in the $(\theta,\phi)$ direction, with modulus $|\vec p_W|$. For the $b$ quark momentum $p_b^{\prime R}$ we do likewise, but in the opposite spatial direction. 
\item The coordinate system in the $W$ rest frame is built as mentioned above, with the $\hat z'$ axis in the direction of $p_W^{\prime R}$ and the $\hat y'$ axis in the $xy$ plane making an angle $\phi$ with the $\hat y$ axis.
\item The new $\ell$ four-momentum in the $W$ rest frame $p_\ell^{\prime R}$ is defined with energy 
$E_\ell$ and three-momentum in the $(\theta^*,\phi^*)$ direction of the above-defined coordinate system, and modulus $|\vec p_\ell |$. For the neutrino momentum $p_\nu^{\prime R}$, the same is done using $E_{\nu}$, $|\vec p_{\nu}|$ and the opposite spatial direction.
\item The new momenta $p_W^\prime$, $p_b^\prime$ in the lab frame are obtained by a two-step boost: first a pure boost of $p_W^{\prime R}$, $p_b^{\prime R}$ from the top rest frame to the c.m. frame, followed by a pure boost from the c.m. frame to the lab frame.
\item The new momenta $p_\ell^\prime$, $p_\nu^\prime$ in the lab frame are obtained by a three-step boost: first a pure boost of $p_\ell^{\prime R}$, $p_\nu^{\prime R}$ from the $W$ rest frame to the top rest frame, followed by a pure boost from the top rest frame to the c.m. frame, and finally a pure boost from the c.m. frame to the lab frame.
\end{enumerate}
Note that the top quark momentum is unchanged, as well as the momenta of any other particles. Therefore, the production kinematics is maintained.

\section{Test: top quark decays}
\label{sec:3}

The soundness of the CAR method to fully reproduce the decay distributions is tested in top quark decays, by comparing the angular distributions obtained in this manner with those obtained using spin projectors at the matrix-element level. We consider top pair production and use the helicity basis~\cite{Bernreuther:2015yna} with three orthogonal vectors $(\hat r,\hat n,\hat k)$ defined as
\begin{itemize}
\item K-axis (helicity): $\hat k$ is a normalised vector in the direction of the top quark three-momentum in the $t \bar t$ rest frame.
\item R-axis: $\hat r$ is in the production plane, defined by $\hat r = \mathrm{sign}(\cos \theta) (\hat p_p - \cos \theta \; \hat k)/\sin \theta$, with $\hat p_p = (0,0,1)$ the direction of one proton in the lab frame, and $\cos \theta = \hat k \cdot \hat p_p$. Because of the $\mathrm{sign}(\cos \theta)$ factor, the definition for $\hat r$ is the same if we use the direction of the other proton $- \hat p_p$.
\item N-axis: $\hat n = \hat k \times \hat r$ is orthogonal to the production plane.
\end{itemize}
For the top anti-quark we keep the same set of axes, therefore the K-axis is the opposite to the helicity axis for the anti-quark.

\begin{figure}[t!]
\begin{center}
\includegraphics[width=7cm,clip=]{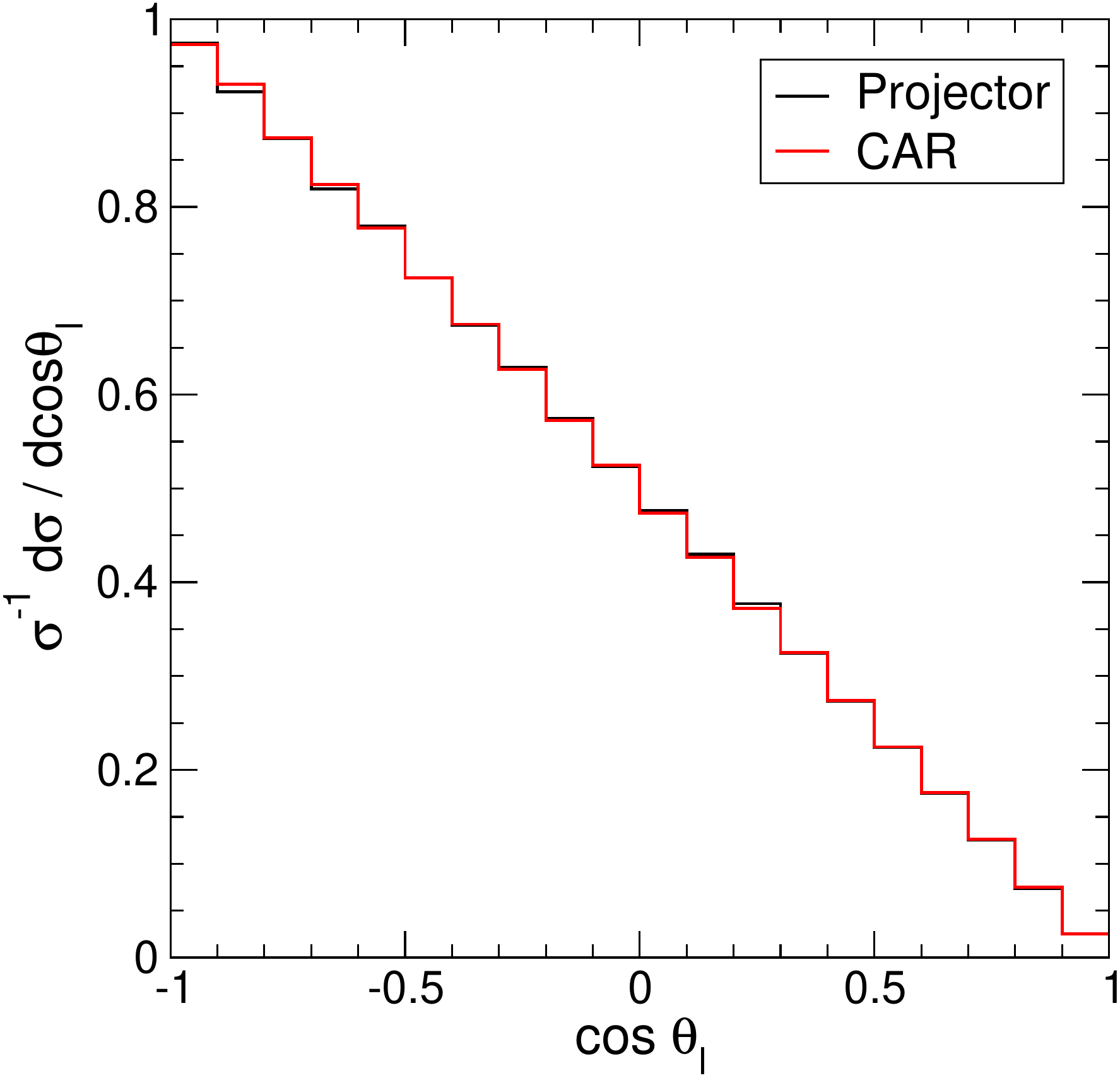}
\caption{Normalised $\cos \theta_\ell$ distributions for top quark samples with $P_3 = -1$ along the K axis (coincident with the $\hat z$ direction), obtained with a spin projector and using the CAR method.}
\label{fig:dist1D}
\end{center}
\end{figure} 

Polarised samples are generated with {\scshape Protos}~\cite{protos} implementing the spin projectors at the matrix-element level~\cite{Aguilar-Saavedra:2021ngj}. Three fully-polarised samples are used, corresponding to $t_L \bar t_R$ using as quantisation axes for both quarks the K, R and N directions above defined. A SM sample is also generated, and processed with the CAR method to obtain $t_L \bar t_R$ samples for the same three quantisation axes K, R and N. In the computation of the amplitude products in (\ref{ec:amp}) we use as $m_t$, $M_W$ the invariant masses computed from the four-momenta, event by event.

\begin{figure*}[t!]
\begin{center}
\begin{tabular}{ccc}
\includegraphics[width=5.5cm,clip=]{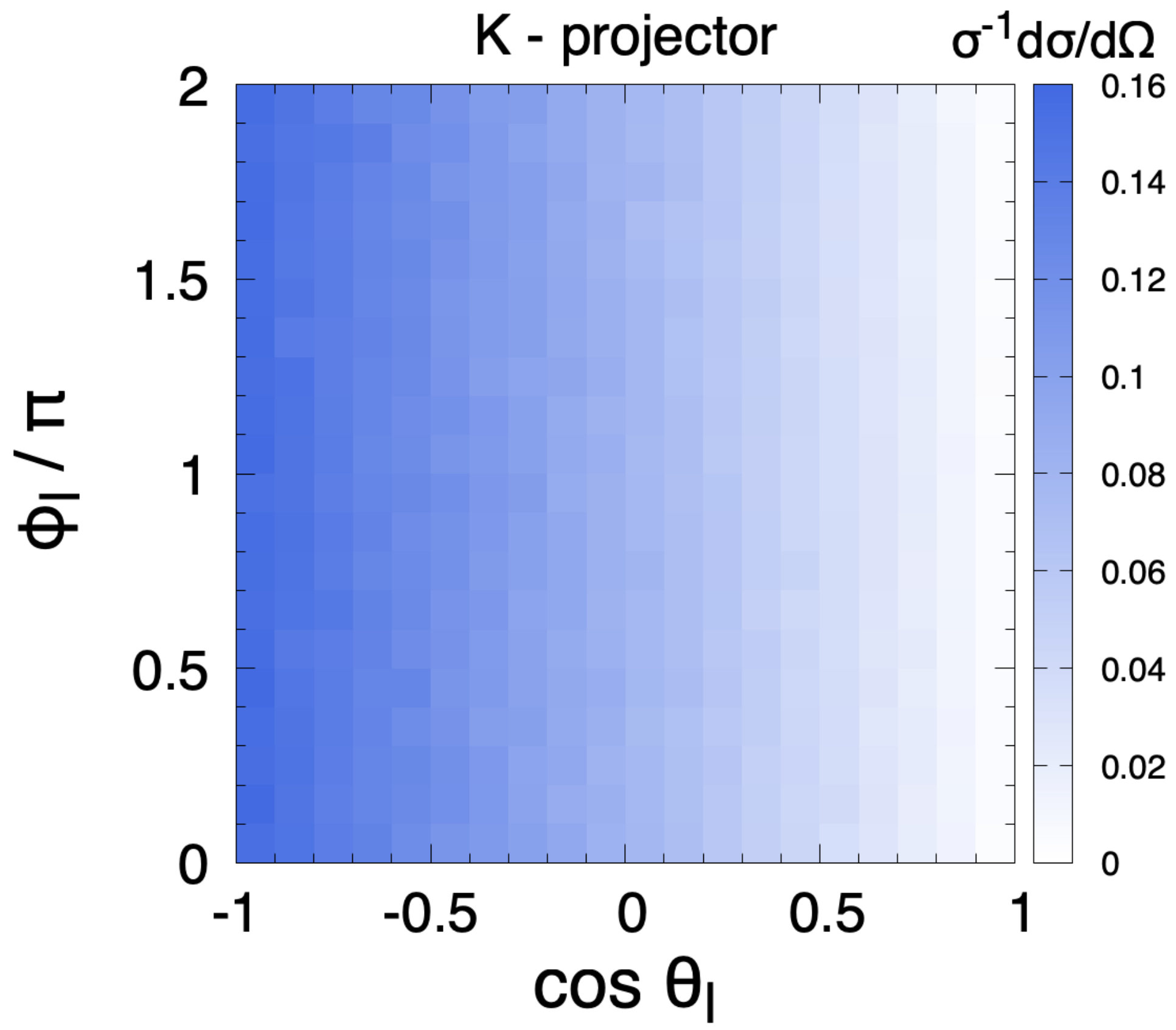} &
\includegraphics[width=5.5cm,clip=]{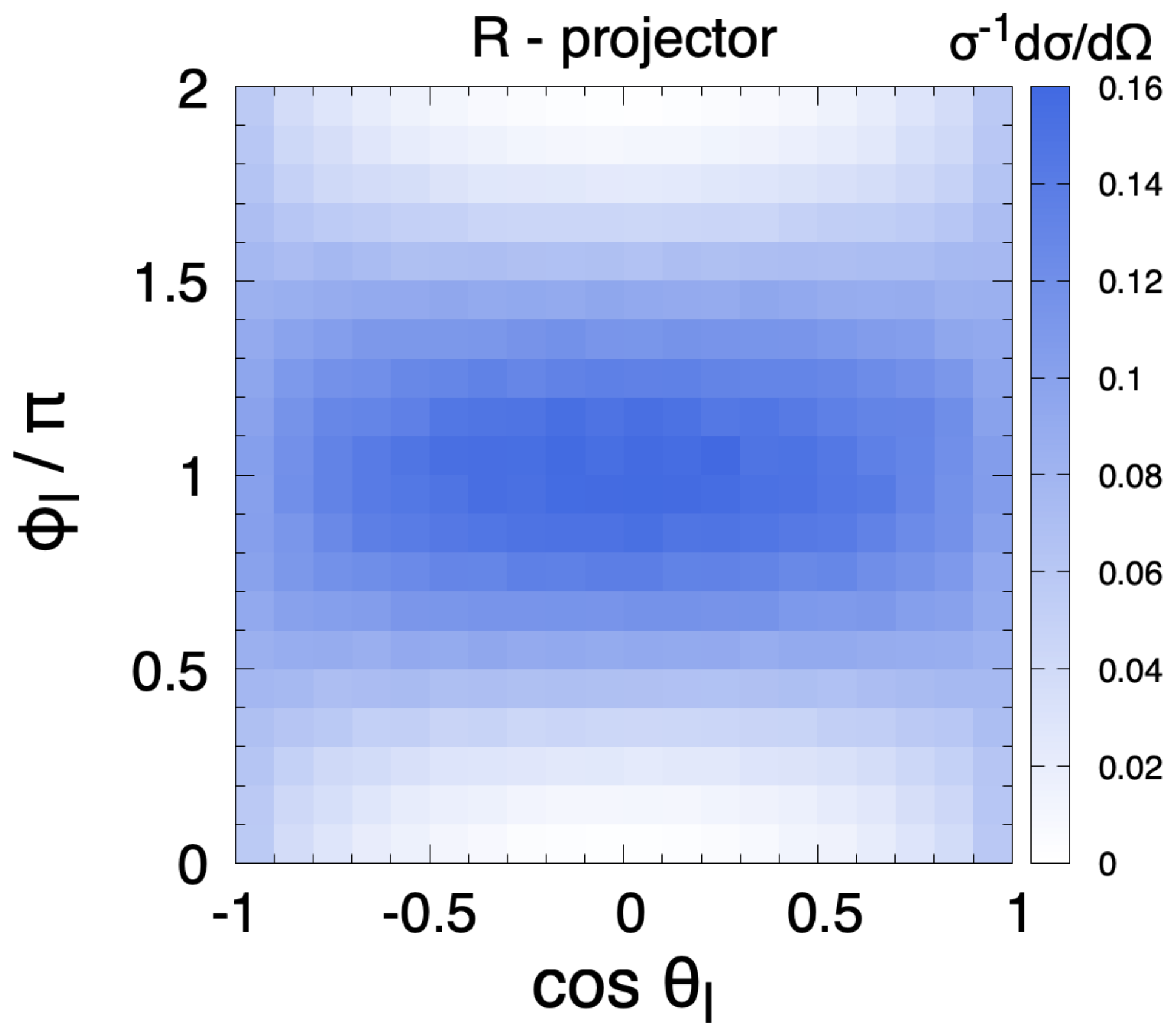} &
\includegraphics[width=5.5cm,clip=]{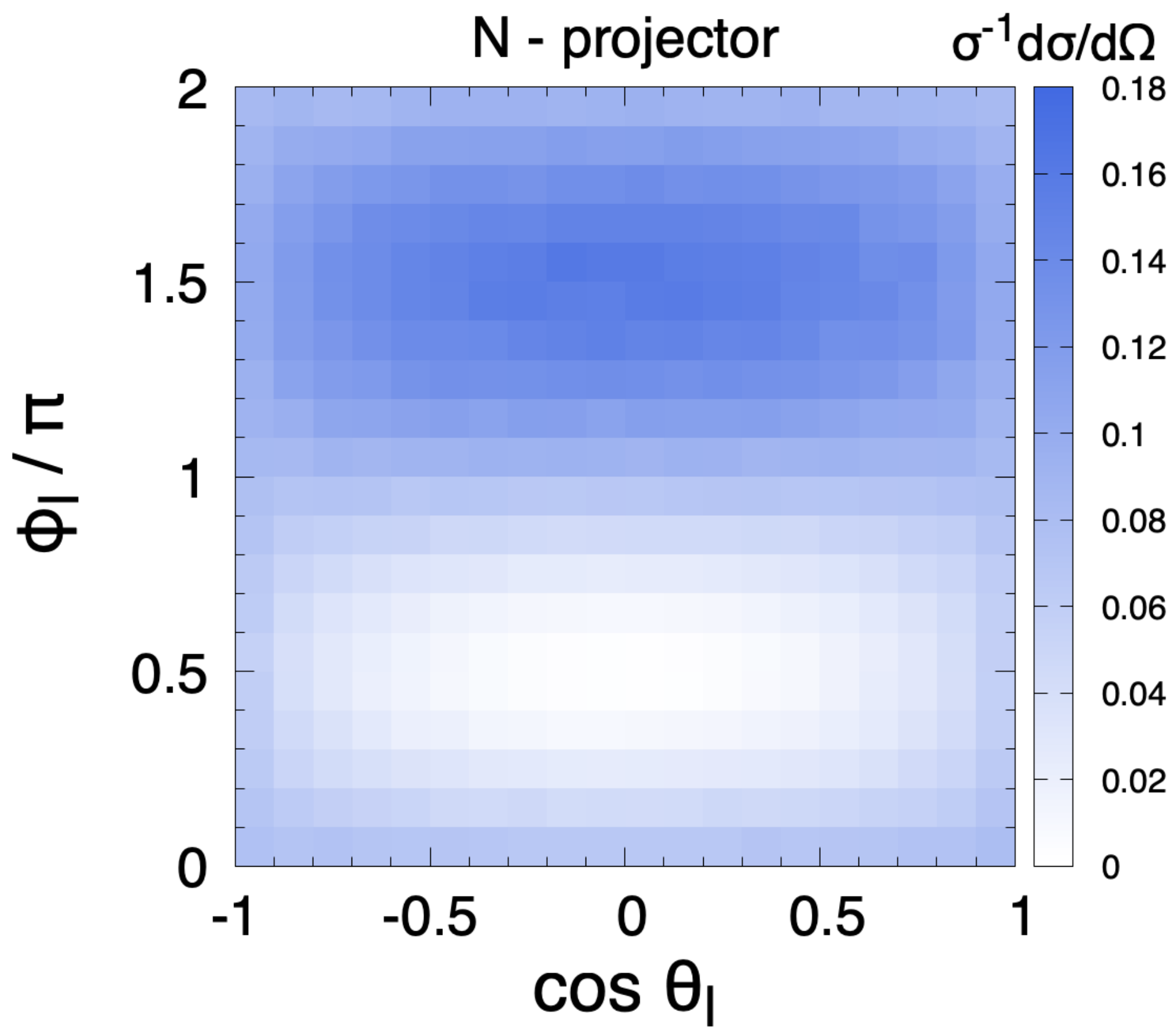} \\
\includegraphics[width=5.5cm,clip=]{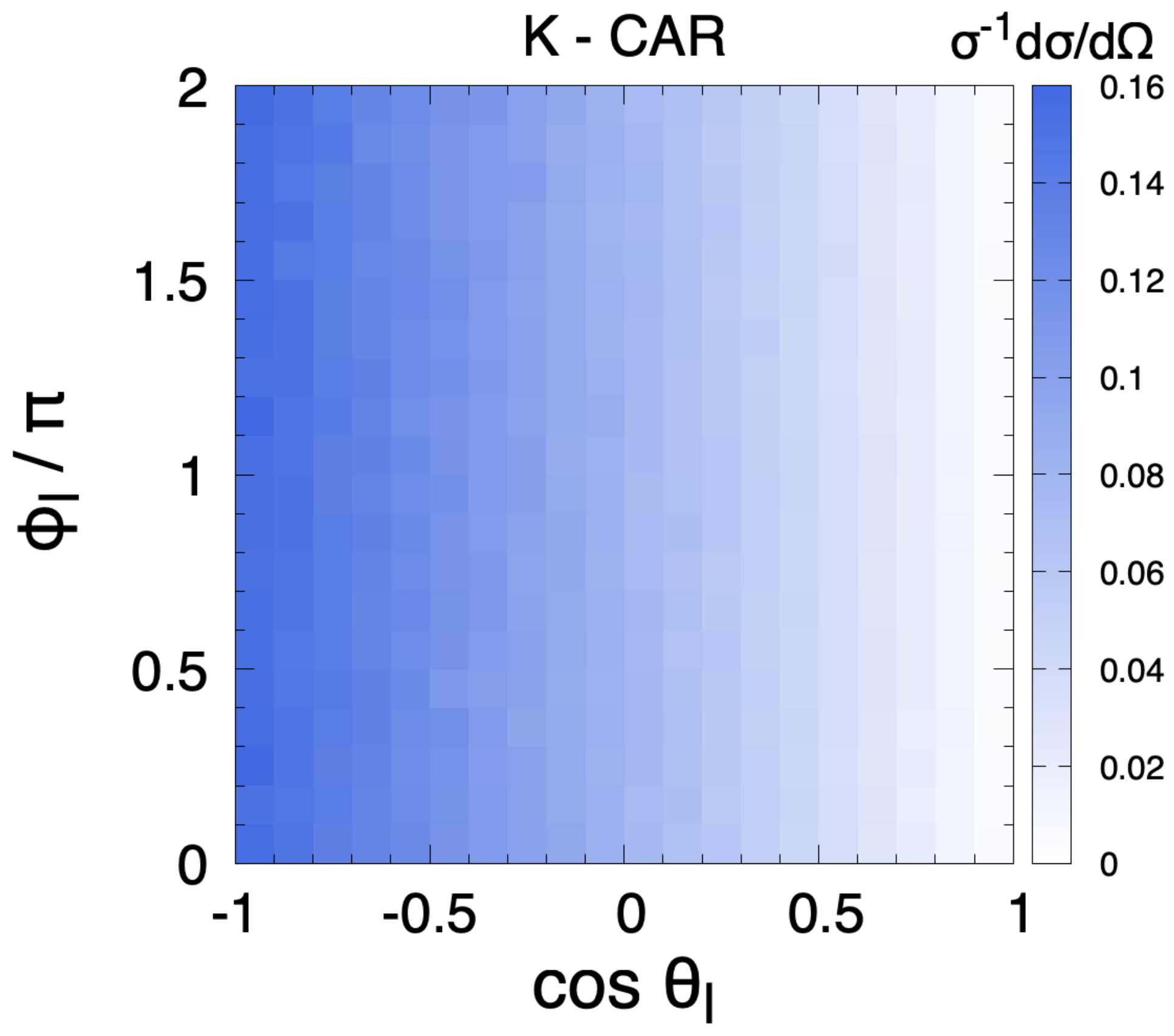} &
\includegraphics[width=5.5cm,clip=]{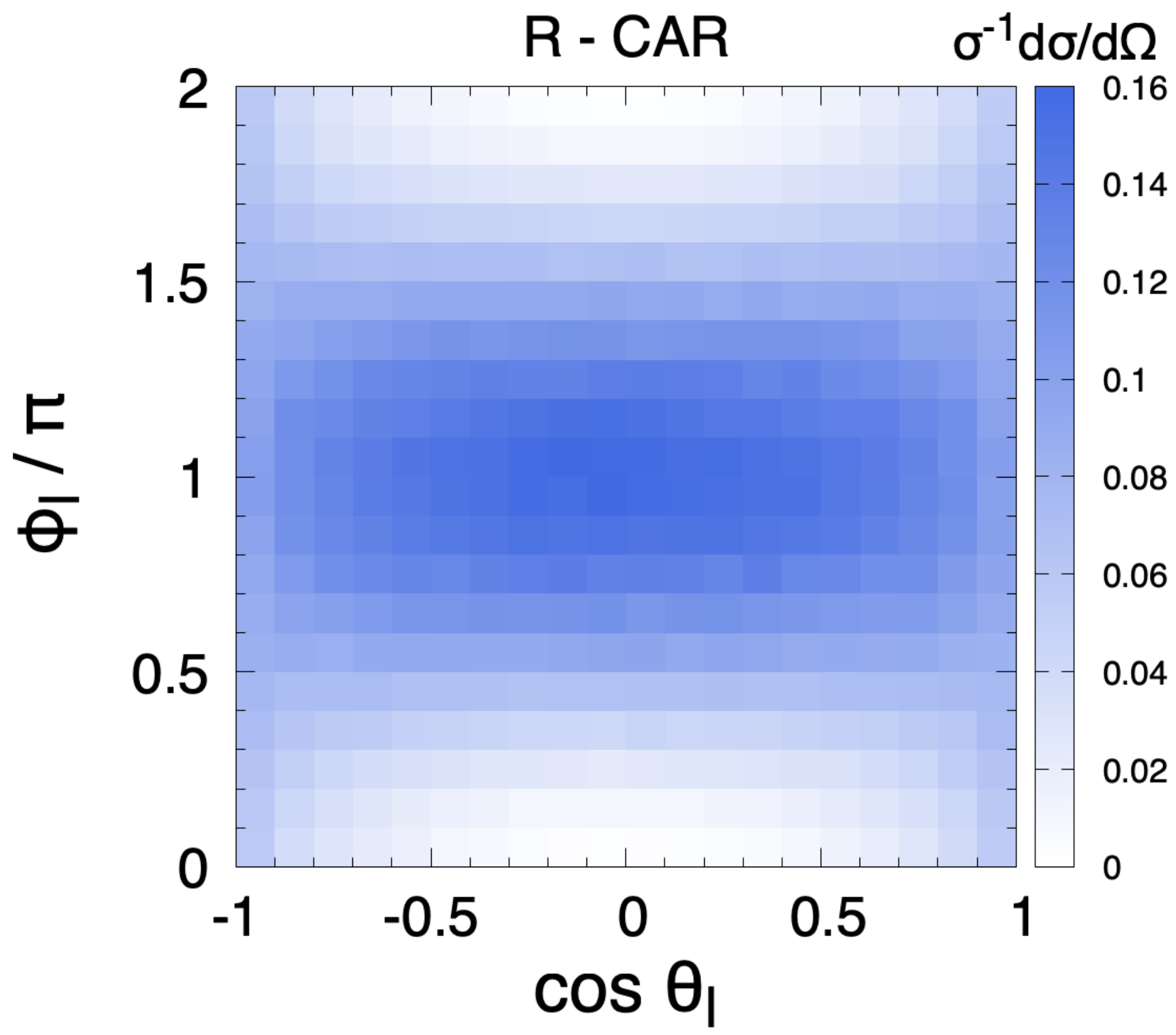} &
\includegraphics[width=5.5cm,clip=]{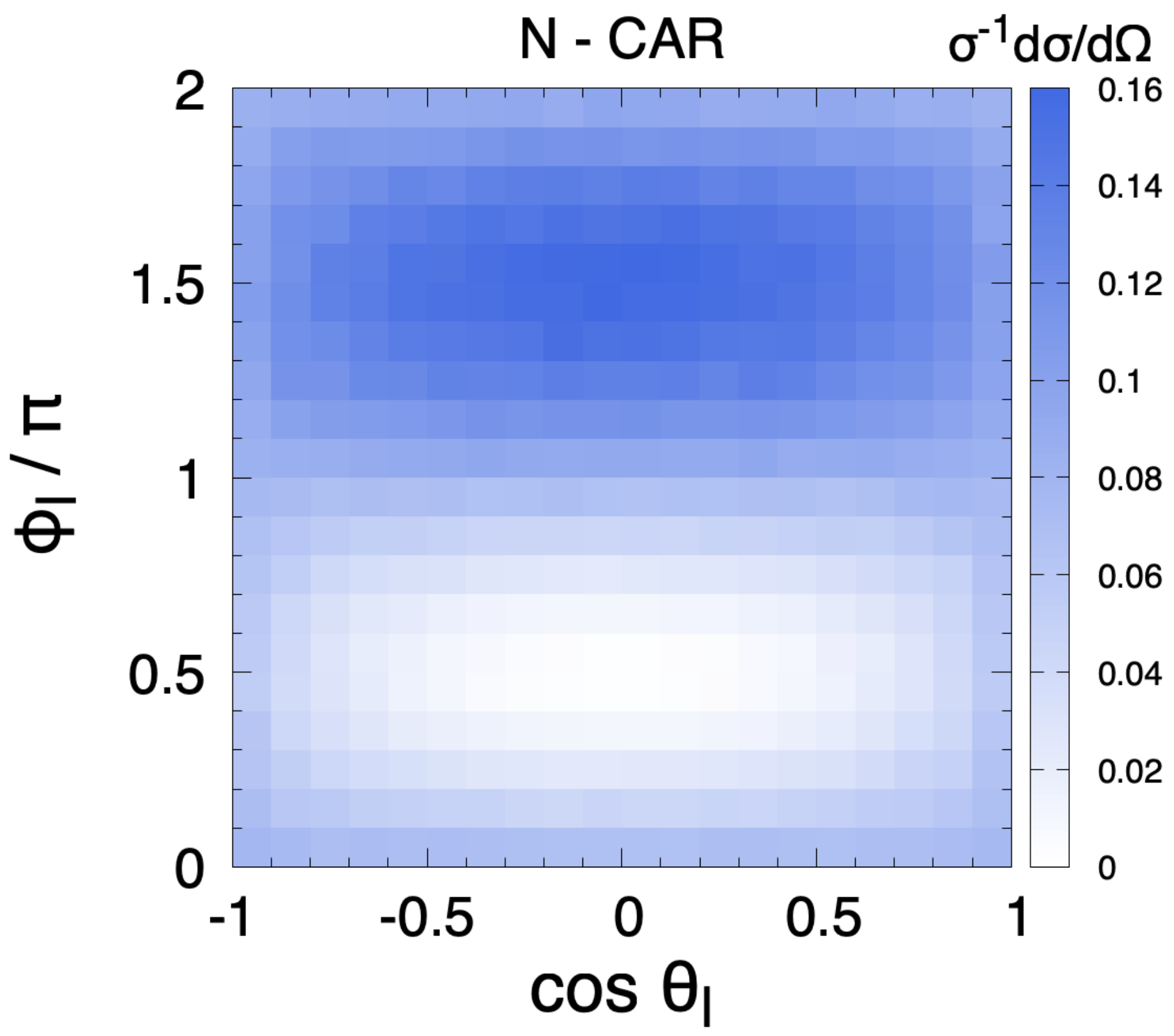} \\
\end{tabular}
\caption{Top: Normalised $(\cos \theta_\ell,\phi_\ell)$ distributions for top quark samples with $P_3 = -1$ along the  K, R and N axes, obtained with spin projectors. Bottom: the same, using the CAR method.}
\label{fig:dist2D}
\end{center}
\end{figure*}

The $\cos \theta$, $\phi$, $\cos \theta^*$, $\phi^*$ one-dimensional distributions are identical (up to statistical fluctuations) for the samples generated with spin projectors or with the CAR method. This is expected because these are precisely the variables used to parameterise the angular dependence of the differential cross section in (\ref{ec:distT}). But a much more stringent test is provided by the charged lepton distribution in the top rest frame. As it is well known, the angle $\theta_\ell$ between the charged lepton $\ell = e,\mu$ in the top rest frame and the top spin direction follows the distribution~\cite{Jezabek:1994zv}
\begin{equation}
\frac{1}{\sigma} \frac{d\sigma}{d\cos\theta_\ell} = \frac{1}{2}(1 + \alpha_\ell \cos \theta_\ell) \,,
\label{ec:distcos}
\end{equation}
with $\alpha_{\ell^+} = - \alpha_{\ell^-} = 1$ in the SM at the LO. This value results from the interference among amplitudes with different $W$ helicities, and then constitutes a crucial non-trivial test of the whole framework. 
The distribution for the positive lepton, using the samples polarised in the K axis, is presented in Fig.~\ref{fig:dist1D}. There is perfect agreement between the sample where the top quark is polarised with a spin projector and the CAR method, and also with the theoretical expectation (\ref{ec:distcos}). The excellent agreement is extensive to the two-dimensional $(\cos \theta_\ell,\phi_\ell)$ distribution. We fix the coordinate axes with $(\hat x, \hat y, \hat z) = (\hat r, \hat n, \hat k)$ and plot in Fig.~\ref{fig:dist2D} the $(\cos \theta_\ell,\phi_\ell)$ distribution for the top quark samples polarised in the negative K, R and N axes, using either a spin projector or the CAR method. The excellent agreement between the two methods shows the validity of the approach.

\section{Application: top pair production}
\label{sec:4}

In this section we investigate the possibility to use the CAR method to augment the size of already polarised data samples. These pre-existing polarised samples are merely used to provide a parameterisation of the production kinematics. And, because the decay kinematics is completely replaced with the CAR method, the resulting samples can be regarded as statistically independent and their use does not introduce any bias, at it will be shown here.

We demonstrate this point with a template fit performed on Monte Carlo pseudo-data, following the approach in Ref.~\cite{Aguilar-Saavedra:2021ngj}, which we briefly review here. We consider top pair production at the LHC, in the dilepton decay mode $t \bar t \to \ell^+ \nu b \ell^- \nu \bar b$. Let us define the short-hand notation $z_+ \equiv \cos \theta_{\ell^+}$, $z_- \equiv \cos \theta_{\ell^-}$, where the angles $\theta_\ell$ have the same definition as in the previous section, and 
\begin{equation}
\bar f(z_+,z_-) = \frac{1}{\bar \sigma} \frac{d \bar \sigma}{dz_+ dz_-}
\end{equation}
the (pseudo-data) normalised distribution that is fitted to extract the $t \bar t$ polarisation coefficients. The bars indicate that the cross sections (integrated and differential) are taken after reconstruction and kinematical cuts. The normalised template distributions for the different polarisation combinations $t_L \bar t_L$, $t_L \bar t_R$, $t_R \bar t_L$, $t_R \bar t_R$ are denoted as
\begin{equation}
\bar f_{XX'}(z_+,z_-) = \frac{1}{\bar \sigma_{XX'}} \frac{d\bar \sigma_{XX'}}{dz_+ dz_-} \,.
\end{equation}
with $X,X' = L,R$. Efficiency factors $\varepsilon = \bar \sigma / \sigma$, $\varepsilon_{XX'} = \bar \sigma_{XX'} / \sigma_{XX'}$ take into account the overall effect of the kinematical cuts in decreasing the parton-level cross sections (without a bar). With this notation, the template expansion reads 
\begin{equation}
\varepsilon \bar f(z_+,z_-) = \sum_{XX'} a_{XX'} \varepsilon_{XX'} \bar f_{XX'}(z_+,z_-) + \Delta_\text{int}(z_+,z_-) \,,
\label{ec:expansion3}
\end{equation}
including a (small) interference term $\Delta_\text{int}$ that cannot be omitted in precision measurements. The polarisation coefficients $a_{XX'}$ are precisely the quantities to be extracted from the fit. The same procedure is repeated taking the $\hat z$ axis in the K, R or N directions, in order to measure the polarisation coefficients and $t \bar t$ spin correlation in these three axes.

For each of the K, R and N axes we generate with {\scshape Protos} two polarised $t_L \bar t_L$ and $t_L \bar t_R$ samples of $5 \times 10^4$ events,  As said, these small samples are meant to parameterise the dependence of the production kinematics (top $p_T$, etc.) on the polarisation. The $t_L \bar t_L$ samples are used to obtain large $t_L \bar t_L$ and $t_R \bar t_R$ samples of $10^6$ events with the CAR method. Likewise, the $t_L \bar t_R$ samples are used to obtain large $t_L \bar t_R$ and $t_R \bar t_L$ samples of  $10^6$ events. For each axis and polarisation combination we generate two statistically-independent samples: one of them is used in the calculation of $\Delta_\text{int}$ (see Ref.~\cite{Aguilar-Saavedra:2021ngj} for details), and the other one for the template fit. Two additional SM samples with $10^6$ events are also generated: one is kept for the $\Delta_\text{int}$ calculation while the other one is used as pseudo-data for the fit.

For simplicity, we do not perform any detector simulation and work at the parton level, but introducing kinematical cuts $p_T \geq 30$ GeV for final-state leptons and $b$ quarks, which are sufficient to strongly modify the SM and template distributions $\bar f(z_+,z_-)$, $\bar f_{XX'}(z_+,z_-)$ with respect to the parton-level ones. The true top momenta are used for the determination of the rest frames and lepton angles. The impact of the kinematical cuts is illustrated by the size of the efficiency factors, collected in Table~\ref{tab:eff}. The size of each sample after the kinematical cuts is $10^6 \times \varepsilon$ events.

\begin{table}[thb]
\begin{center}
\begin{tabular}{cc}
Sample & $\varepsilon$ \\[1mm]
SM & $0.362$
\end{tabular}

\vspace{3mm}

\begin{tabular}{cccc}
Template & \multicolumn{3}{c}{$\varepsilon$} \\
     & K-axis & R-axis & N-axis \\[1mm]
$LL$ & $0.288$ & $0.433$ & $0.367$ \\
$RR$ & $0.347$ & $0.349$ & $0.364$ \\
$LR$ & $0.346$ & $0.350$ & $0.364$ \\
$RL$ & $0.508$ & $0.329$ & $0.368$
\end{tabular}
\caption{Efficiencies for the samples and templates considered in this work. The uncertainties are of the order of $10^{-3}$.} 
\label{tab:eff}
\end{center}
\end{table}

Table~\ref{tab:coeffs} collects the true value of the $t \bar t$ polarisation coefficients for the SM sample and the values obtained from the template fit using the samples with kinematical cuts. The Monte Carlo statistical uncertainties expected for samples with $2.9 \times 10^5 - 10^6$ events are of the order of $10^{-3}$, as it can also be verified by comparing coefficients that are equal at the leading order, namely $a_{LL} = a_{RR}$, $a_{LR} = a_{RL}$. The difference between the true and fitted values for the coefficients is also of the order of $10^{-3}$, showing that there is no bias in the polarisation templates obtained with the CAR method. The last column of Table~\ref{tab:coeffs} shows the spin correlation coefficients $C = a_{LL} + a_{RR} - a_{LR} - a_{RL}$. The Monte Carlo statistical uncertainties on spin-correlation coefficients are of the order of $10^{-2}$.

\begin{table}[htb]
\begin{center}
\begin{tabular}{ccccccc}
& & $a_{LL}$ & $a_{LR}$ & $a_{RL}$ & $a_{RR}$ & $C$ \\[0.5mm]
\multirow{2}{*}{K-axis} & True & 0.164 & 0.337 & 0.337 & 0.161 & -0.349 \\[0.5mm]
                                    & Fit    & 0.168 & 0.333 & 0.335 & 0.165 & -0.334 \\[0.5mm]
\multirow{2}{*}{R-axis} & True & 0.244 & 0.256 & 0.254 & 0.245 & -0.021 \\[0.5mm]
                                    & Fit    & 0.243 & 0.259 & 0.257 & 0.242 & -0.031 \\[0.5mm]
\multirow{2}{*}{N-axis} & True & 0.168 & 0.333 & 0.333 & 0.169 & -0.331 \\[0.5mm]
                                    & Fit    & 0.165 & 0.333 & 0.335 & 0.166 & -0.336
                                    
\end{tabular}
\end{center}
\caption{True values (extracted from the Monte Carlo sample without kinematical cuts) best-fit values for various polarisation coefficients for the SM sample. The uncertainties on the $a$ coefficients are of the order of $10^{-3}$, and on $C$ coefficients of the order of $10^{-2}$.}
\label{tab:coeffs}
\end{table}

We finally test how important the polarisation effects in the $t \bar t$ production kinematics are. We repeat the fit but this time using as templates SM samples where only the top (anti-)quark decays are modified. The results for the polarisation coefficients are given in Table~\ref{tab:coeffs2}. While for the N axis there is no difference at this level, for the K and R axes there are significant deviations from the correct values in Table~\ref{tab:coeffs}. This example shows that, for this type of measurement, polarised samples must include the effect of polarisation both in the production and the decay. This is understood because the $p_T$ of the top quark decay products depend on the top quark kinematics ($p_T$ and rapidity) as well as on the decay angular distributions. A $p_T$ cut modifies the templates in a way that depends on the top kinematics itself, which in turn depends on the polarisation.

\begin{table}[t]
\begin{center}
\begin{tabular}{cccccc}
& $a_{LL}$ & $a_{LR}$ & $a_{RL}$ & $a_{RR}$ & $C$ \\[0.5mm]
K-axis & 0.176 & 0.322 & 0.323 & 0.179 & -0.290 \\[0.5mm]
R-axis & 0.255 & 0.244 & 0.242 & 0.259 & 0.028 \\[0.5mm]
N-axis & 0.167 & 0.332 & 0.333 & 0.168 & -0.330
\end{tabular}
\end{center}
\caption{Value of the polarisation coefficients obtained using templates that keep the SM unpolarised production kinematics. The uncertainties on the $a$ coefficients are of the order of $10^{-3}$, and on $C$ coefficients of the order of $10^{-2}$.}
\label{tab:coeffs2}
\end{table}

\section{Discussion}
\label{sec:5}

The method introduced here allows to tune the polarisation of heavy particles in a pre-existing Monte Carlo event sample. This `portability' makes it very useful: an event sample can be generated with any code and subsequently processed with the CAR method to tune the polarisation of the desired particles. As mentioned, this can be done even if the sample is generated beyond the LO in production, though the method can only be used within the production $\times$ decay approximation. The validity of the approach is demonstrated in section~\ref{sec:3} with an example for the top quark decay $t \to W b \to \ell \nu b$. For longer cascade decay chains the procedure would be exactly as outlined here, but with more intermediate steps and additional definitions for reference systems. Note also that a similar goal might be achieved by reweighting the decay distributions according to (\ref{ec:distV}) or (\ref{ec:distT}); however, {\em replacing} as in the CAR method is computationally more efficient than {\em reweighting} since the former keeps all events with unit weight and does not entail an effective loss of Monte Carlo statistics.

By construction, the CAR method keeps the production kinematics of the pre-existing sample, only modifying the decay of the $t$, $W$ or $Z$ particles. In some cases this may constitute an advantage, e.g. to test the robustness of unfolding methods on polarised samples keeping the SM kinematics. But some applications require modified kinematics as well. For example, for top polarisation measurements using a template method as in Ref.~\cite{Aguilar-Saavedra:2021ngj}, the templates must be generated using spin projectors at the matrix-element level, so that the correct polarisation dependence --- consistent with the definition of spin for an intermediate resonance --- is kept both in the top (anti-)quark production and the decay distributions~\cite{Aguilar-Saavedra:2022jzo}.
The CAR method can still be very useful in this case. Since the phase space for the production has a lower dimensionality, it is sufficient to generate a small polarised sample with the appropriate production kinematics, using spin projectors. This sample can be subsequently processed with the CAR method to obtain quite large samples of events that are (almost) statistically independent. This sample augmentation can bring a computational advantage for processes with multiple top quarks and/or weak bosons, especially for event generation beyond the LO. 

Finally, let us comment that in the analytical expressions collected in this work (see section~\ref{sec:2}), the decay of the top quark and $W/Z$ bosons are implemented at the LO. This does not seem to entail a serious limitation, as the experimentally most interesting channels involve (semi-)leptonic decays where next-to-leading order (NLO) corrections are small. For example, NLO corrections to $\alpha_\ell$ in (\ref{ec:distcos}) are at the permille level~\cite{Brandenburg:2002xr}. In any case, calculations beyond the LO, e.g. for the top quark decay~\cite{Fischer:2001gp}, exist in the literature and might eventually be implemented if the experimental precision requires it.

\section*{Acknowledgements}

I thank M.L. Mangano and J. Alcaraz for very useful discussions, and the CERN Theory Department for hospitality during the completion of this work.
This work has been supported by 
 the grants IFT Centro de Excelencia Severo Ochoa 
CEX2020-001007-S and PID2019-110058GB-C21, funded by MCIN/AEI/10.13039/501100011033 and by ERDF, 
and by FCT project CERN/FIS-PAR/0004/2019. 

\bibliographystyle{utphys}
\bibliography{references}

\end{document}